\documentclass[fleqn,twoside]{article}
\usepackage{espcrc2}
\usepackage{psfig}

\newcommand{\AmS}{{\protect\the\textfont2
  A\kern-.1667em\lower.5ex\hbox{M}\kern-.125emS}}

\newcommand\chandra{{\em Chandra}}

\newcommand\xmm{{\em XMM-Newton}}

\newcommand\casa{Cas~A}
\newcommand\msh{G292.0+1.8}

\newcommand\msun{{M$_{\odot}$}}
\newcommand\net{$n_{\rm e}t$}

\newcommand{\Te}{{$kT_{\rm e}$}}
\newcommand{\NH}{{$N_{\rm H}$}}

\newcommand{\netunit}{{cm$^{-3}$s}}
\newcommand{\kms}{{km\,s$^{-1}$}}
\newcommand\arcmin{\mbox{$^\prime$}}%

\def\apj{{ApJ\,}}
\def\apjl{{ApJL\,}}

\def\aap{{A\&A\,}}

\title{High Resolution X-ray Spectroscopy of G292.0+1.8/MSH 11-5{\it 4}}

\author{Jacco Vink,\thanks{Chandra fellow}
\address[columbia]{Columbia Astrophysics Laboratory, Columbia University, 
New York, NY, USA}
\address[sron]{SRON National Institute for Space Research,
Sorbonnelaan 2, 3584 CA, Utrecht, Netherlands}
Johan Bleeker\addressmark[sron], Jelle S. Kaastra\addressmark[sron], 
Andrew Rasmussen\addressmark[columbia]}

\begin{document}

\begin{abstract}
We present a preliminary analysis of \xmm\ observations of
the oxygen-rich supernova remnant G292.0+1.8 (MSH\,11-5{\it 4}). 
Although the spatial extent of the remnant is 8\arcmin\ the bright
central bar is narrow (1-2\arcmin) resulting
in RGS spectra of a high spectral quality.
This allows us to spectroscopically identify a cool, \Te $= 0.3$~keV,
and underionized component, resolve details of
the Fe-L complex, and resolve the
forbidden and resonant lines of the O\,VII triplet.
We are also able to constrain the kinematics of the remnant using
Ne\,IX as observed in the second order spectrum, and O\,VIII in
the first order spectrum.
We do not find evidence for O\,VII line shifts or
Doppler broadening 
($\sigma_v < 731$~\kms), but line broadening of the Ne\,X Ly$\alpha$\
line seems to be present, corresponding to $\sigma_v \sim 1500$~\kms.
\end{abstract}

\maketitle

\section{INTRODUCTION}
\msh\ (MSH\,11-5{\it 4}) belongs to  the class of
oxygen-rich supernova remnants,
which are probably the products of core collapse supernovae of the 
most massive stars, i.e. a main sequence mass in excess of $\sim 20$\msun.\footnote{See the review by Vink, these proceedings.}
It has a size of $\sim8$\arcmin\, and its morphology is characterized by
a filamentary X-ray structure, and 
a bar stretching from the central east to the west side.
The nature of this bar is unclear, but may have something to do with
a structure in the progenitor's circumstellar medium \cite{park02}.

\begin{figure}
\psfig{figure=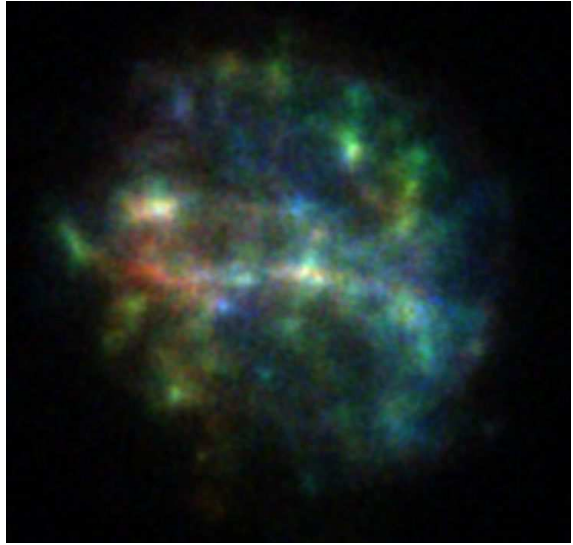,width=\columnwidth}
\caption{\xmm\ EPIC-MOS image of \msh. 
The RGB colors code for the line emission in O\,VIII,
Mg XI, and Si XIII.
\label{image}}
\end{figure}

Recently it was discovered that a region of hard X-ray emission, east of
the center, harbors a 135~ms radio/X-ray pulsar \cite{camillo02,hughes03b}.
This is of special interest as it is likely that \msh\ is the remnant
of a very massive star (30-40~\msun, \cite{gonzalez03}). Such massive
stars are thought to give birth to black holes \cite{heger03}.
Surprisingly, there are only a few examples of shell type supernova remnants
with pulsar wind nebulae inside.

\begin{figure*}
\centerline{
\psfig{figure=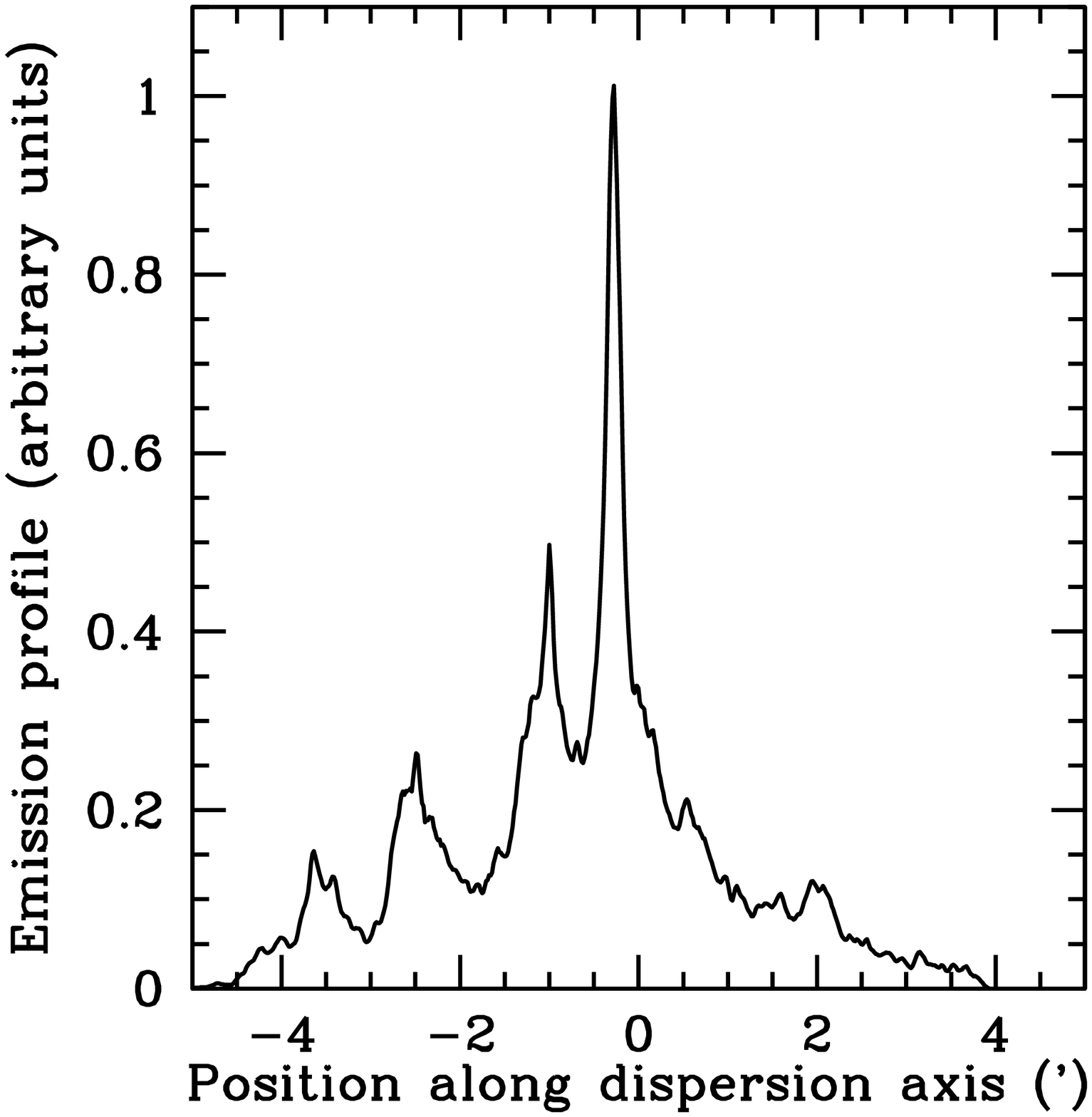,width=0.4\textwidth}
\psfig{figure=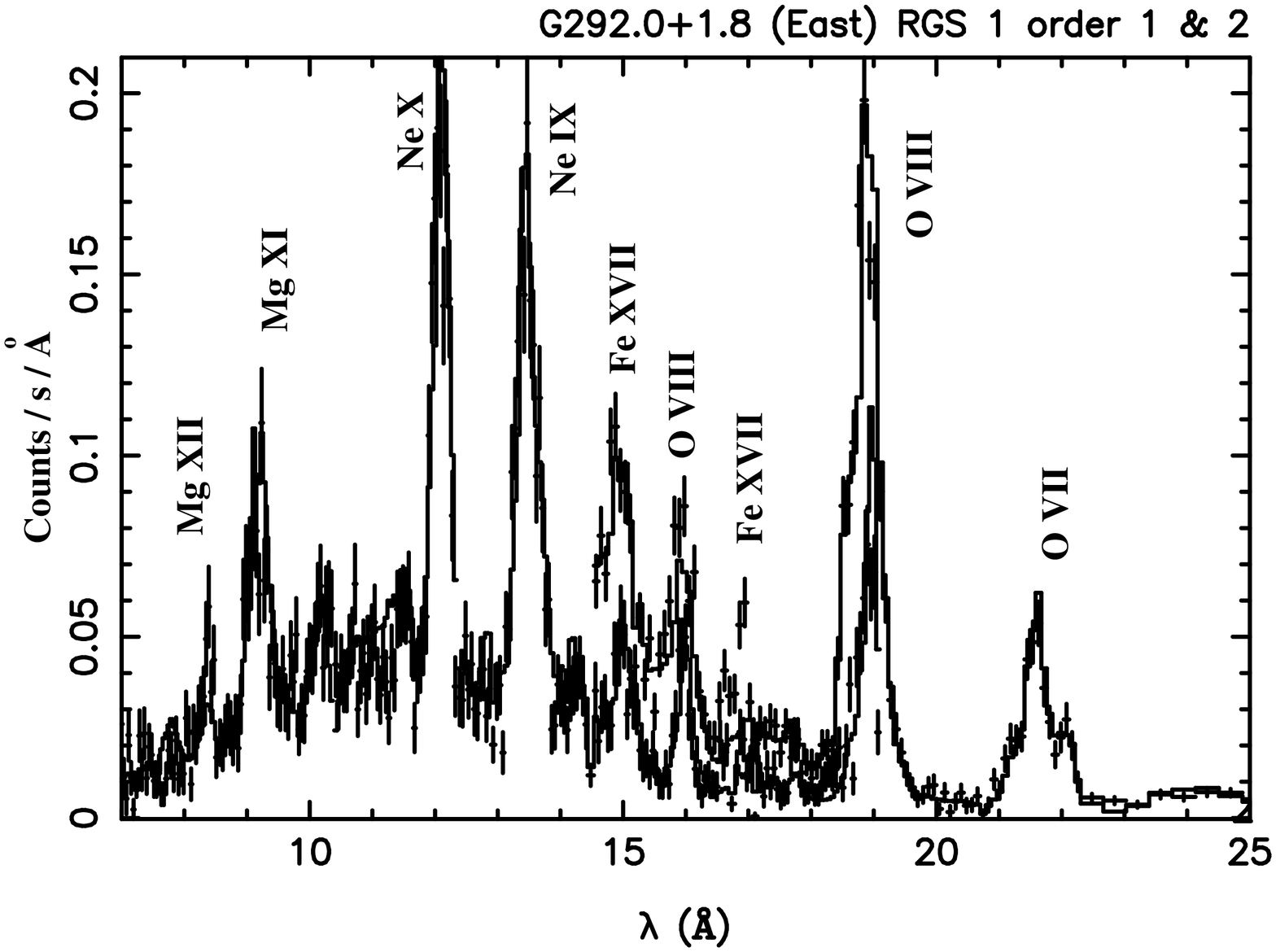,width=0.55\textwidth}
}
\caption{ X-ray emission profile based on the high spatial resolution
\chandra\ images, in the direction of the \xmm\ RGS dispersion axis,
using the O\,VII He$\alpha$\ band (left).
The sharp peak in the profile results in a high resolution
\xmm\ RGS spectrum (right). 
\label{spec}}
\end{figure*}

\chandra\ observations of \msh\ 
have revealed substantial variations in abundances
within the remnant \cite{park02,gonzalez03}, the most prominent
X-ray lines being O, Ne, Mg, Si and S.

\xmm\ observed \msh\ in August 2002, with a total observation
time of $\sim$28~ks.
Here we report on a preliminary analysis of the data
with an emphasis on the reflection grating spectrometer 
(RGS, \cite{denherder01}) data.

\section{HIGH RESOLUTION SPECTROSCOPY}
The peculiar bar-like feature across the remnant has an advantage
for observation by the RGS instrument, as long
as the dispersion axis is perpendicular to the bar,
as was the case during this observation. 
As the RGS is a slitless
spectrometer, the spectral resolution is degraded by the spatial
extent of the observed object. The degradation is approximately
0.12~\AA\ per arcmin. However, the bar is bright with respect
to the rest of the remnant and has a width of 1-2\arcmin\ (Fig.~\ref{spec}).
The effective spectral resolution in FWHM is therefore $\sim2$~\AA.
The relative spectral resolution increases at longer wavelengths. For
shorter wavelengths higher resolution can be obtained by extracting
second order spectra.

In order to perform a quantitative analysis of the RGS spectra the
response matrix has to be convolved with the spatial profile
of the remnant (see \cite{rasmussen01,vink03b}), 
for which we used the \chandra-ACIS images, which
have a higher spatial resolution. The \chandra\ images were extracted
in narrow energy bands in order to obtain the right profile for
each element.

As can be seen in Fig.~\ref{spec}, the RGS spectrum reveals line features
which are blended when observed with the \chandra\ or \xmm\ CCD detectors
(c.f. the spectra in \cite{park02}). The 
Fe XVII lines are of interest as they are weak and
have not been resolved previously.

The spectral resolution is of sufficient quality
in order to estimate the
forbidden over resonant line ratios of the O\,VII He$\alpha$ triplet,
which is an important plasma diagnostic \cite{porquet01}.
The best fit with  an absorption of \NH $ = 4\times10^{21}$~cm$^{-2}$ 
gives a G-ratio of $(f+i/r)  = 0.46-0.91$ (Fig.~\ref{triplet}).
According to the SPEX non-equilibrium ionization model,
the triplet emission indicates
a range for \Te\ of 0.27 - 0.51~keV (90\% confidence).

\section{A COOL COMPONENT}
\begin{figure}
\psfig{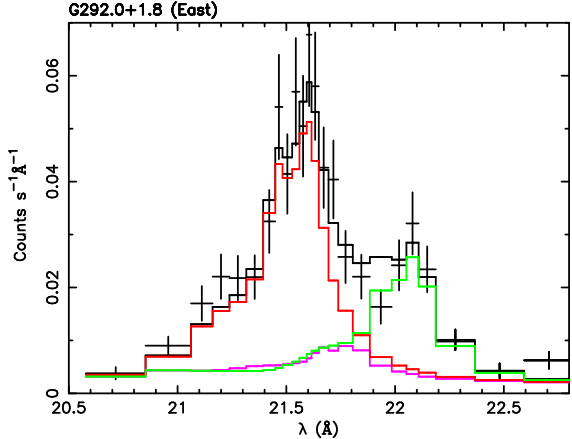}
\caption{
Detail of RGS 1 order 1 spectrum of the eastern region. It shows the 
decomposition of the O\,VII triplet in resonance (red), intercombination (magenta), and forbidden line emission (green). 
\label{triplet}}
\end{figure}

Previous X-ray studies of \msh\ typically found that the plasma
temperature is \Te = 0.7~keV, 
although an additional low 0.3~keV temperature component was found 
in a region near the rim of \msh\ \cite{gonzalez03}. 
However, fitting the RGS spectra with the 
SPEX non-equilibrium ionization model
indicates that in addition to the 0.7~keV plasma component
a much cooler
component with \Te $\sim 0.3$~keV has to be present throughout the remnant
(see also the previous section).
The spectral fits indicate that this component is also
considerably more underionized than the 0.7~keV component:
\net $\le 4\times10^{10}$~\netunit\ versus \net  $\sim 10^{11}$~\netunit.
The cool component varies in importance, with the
contribution of the cool component to the total emission
measure varying from 13\% (in the west) to 28\% 
(east).
This component is the main source of the
O\,VII emission shown in Fig.~\ref{spec} and Fig.~\ref{triplet}

\section{THE KINEMATICS OF \msh}
In order to get a better idea of the ejected mass, the explosion energy,
and the evolutionary phase of the \msh, kinematical data are
important, 
as the example of \casa\ shows \cite{vink98a,willingale02,delaney03}.  
We are currently working on modeling the line profiles,
and directly on the dispersed images in order to obtain kinematical
information. This work is still in progress, but our preliminary results
indicate that the O\,VII Ly$\alpha$ is coming from a plasma with no
great bulk motions ($v  < 114$~\kms) and no apparent Doppler line
broadening ($\sigma_v< 730$~\kms, 95\% confidence level).
However, the Ne\,X Ly$\alpha$\ emission, measured using the
2nd order spectra, is best fitted 
by including line broadening corresponding to
$\sigma_v = 880  - 2880$~\kms (95\% confidence range).
Previous studies indicated that most of the Ne is associated with
the ejecta, whereas the O emission has a large contribution from
shock heated interstellar/circumstellar material.
Our results therefore suggests that the blastwave has decelerated considerably,
whereas at least some of the ejecta, presumably the material
in relatively dense knots, are still moving with a high velocity.

\section{CONCLUSIONS}
We have presented preliminary results of an analysis
of \xmm\ observations of \msh. We hope to improve on this
analysis in the near future.
The tentative conclusions are:
\begin{itemize}
\itemsep -0.1cm
\item It is possible to obtain 
high resolution RGS spectra from a relatively large object
as \msh, provided that narrow, outstanding features exist in
the spatial emission profile along the dispersion axis.
\item The RGS spectra of \msh\ indicate the presence of at least
two temperature components, with \Te $\sim$ 0.7 keV and, 
somewhat unexpectedly, 
\Te $\sim$ 0.3 keV.
The coolest component is needed in order to account for the
observed O\,VII line emission.
\item No significant line broadening  (temperature of the  shell) is
  indicated by the O\,VIII Ly$\alpha$ (0.65 keV) line emission: 
  $\sigma_E<1.4$~eV  \hbox{($<731$~\kms)} (95\% confidence),
but significant line broadening seems to be present for the Ne\,X  Ly$\alpha$
emission, corresponding to a velocity dispersion of 
$\sigma_v \sim 1500$~\kms.
\end{itemize}

\vskip 0.1cm
This work was supported by NASA's
Chandra Postdoctoral Fellowship Award Nr. PF0-10011
issued by the Chandra X-ray Observatory Center, which is operated by the
SAO under NASA contract NAS8-39073.
This work is based on observations obtained with \xmm,
an ESA science mission, funded by its member states and the NASA
(USA). SRON is financially supported by NWO, the Netherlands Organization
for Scientific Research.

\bibliographystyle{h-elsevier3}


\end{document}